\documentclass[prb,twocolumn,showpacs,preprintnumbers,amsmath,amssymb,superscriptaddress]{revtex4}

\usepackage{epsfig}
\usepackage{hyperref}

\begin{document}

\title{Enumeration of many-body skeleton diagrams}

\author{Luca Guido Molinari}
\affiliation{Dipartimento di Fisica
dell'Universit\`a degli Studi di Milano, Via Celoria 16, I-20133 Milano\\
and European Theoretical Spectroscopy Facility (E.T.S.F.)}
\affiliation{I.N.F.N. Sezione di Milano}
\author{Nicola Manini}
\affiliation{Dipartimento di Fisica
dell'Universit\`a degli Studi di Milano, Via Celoria 16, I-20133 Milano\\
and European Theoretical Spectroscopy Facility (E.T.S.F.)}
\affiliation{INFM-CNR, Unit\`a di Milano, Milano, Italy}

\date{\today}

\begin{abstract}
The many-body dynamics of interacting electrons in condensed matter and
quantum chemistry is often studied at the quasiparticle level, where the
perturbative diagrammatic series is partially resummed.
Based on Hedin's equations for self-energy, polarization, propagator,
effective potential, and vertex function in zero dimension of space-time,
dressed Feynman (skeleton) diagrams are enumerated.
Such diagram counts provide useful basic checks for extensions of the
theory for future realistic simulations.
\end{abstract}


\maketitle

\section{Introduction}
Current research in electronic properties of molecules, nanocomponents, and
solids has gone far beyond the mean-field density-functional description.
Many-body methods are employed routinely, at the level of quasiparticles,
to describe excitations accurately\cite{Oni,Aul,Arya}.
Propagators and interactions are renormalized to obtain an effective theory
of weakly interacting quasiparticles.
This is achieved by infinite resummations of Feynman diagrams, that are
brought to a smaller class of skeleton diagrams\cite{Bjorken}.
The reduction in number should correspond to better analytic behavior in
space-time of the individual dressed diagrams, as it occurs in random-phase
or ladder resummations \cite{Fetter}.

\begin{figure}
\centerline{\epsfig{file=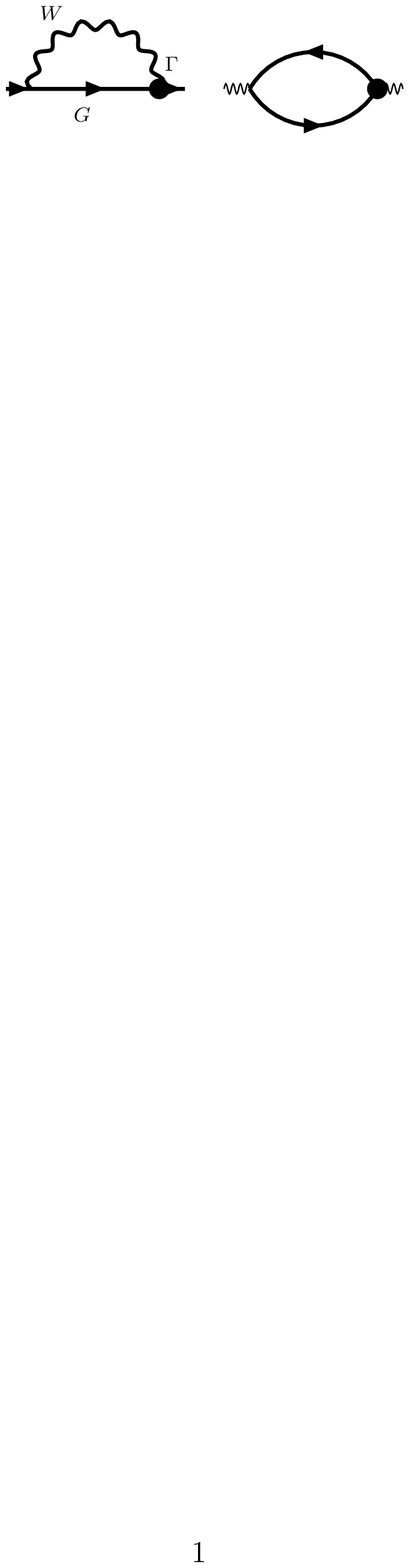,width=85mm,clip=}}
\caption{\label{sigmapi_dr:fig}
The unique fully dressed skeleton diagrams for self-energy (left) and
polarization (right).}
\end{figure}

In this report, we show how skeleton Feynman diagrams of various types can
be enumerated for the exact many-body theory and its GW approximation
\cite{Aul,Arya,Oni}.
We take as our starting point the standard formulation of the many-body
problem of $N$ interacting fermions provided by the set of five Hedin's
equations \cite{Hedin,Hedinlund,Aul,Arya} for the propagator $G(1,2)$,
effective potential $W(1,2)$, irreducible self-energy $\Sigma (1,2)$,
irreducible polarization $\Pi(1,2)$, and irreducible vertex
$\Gamma(1;2,3)$.

In zero dimension of space {\em and} time, the combinatorial content of
Wick's expansion survives, producing the same Feynman diagrams in the
perturbative expansion of the correlators as in conventional 3+1
dimensions.
These correlators no longer carry space-time labels, and do not correspond
to space-time functions.
However, they continue to solve Hedin's equations, which can indeed be
derived from considerations about the topology of diagrams
\cite{Bjorken}.
In zero dimensions ($d=0$), four of Hedin's equations become algebraic, in
terms of the scalar variables $g$ and $v$,
representing the Hartree propagator (computed with the exact density
provided by $G$) and the bare interparticle interaction respectively.
The functional derivative in the equation for the vertex becomes an
ordinary derivative.
The five equations read:
\begin{eqnarray}
G&=g+g\Sigma G\,,\qquad & W=v+v\Pi W\,, \label{HedinGW}\\
\Sigma&=iGW\Gamma\,,\qquad\ \ &  \Pi= i\ell G^2\Gamma \label{Hedinsigmapi}
\,,\\
\Gamma&=1+\Gamma\,G^2\,\frac{\partial \Sigma}{\partial G} \label{Hedinvertex} 
\,.\ &
\end{eqnarray}
The parameter $\ell $ is introduced to count fermion loops: in Hedin's
equations for the electron, $\ell=-2$ because of spin degeneracy and
Feynman's rule for the fermionic loop.
The diagrammatic meaning of Hedin's equations is simple:
Eqs.~(\ref{HedinGW}) correspond to Dyson's equations that define the proper
self-energy and polarization insertions for the propagator and the
effective potential, Eqs.~(\ref{Hedinsigmapi}) translate the unique
skeleton structure\cite{Fetter} of $\Sigma $ and $\Pi $, shown in
Fig.~\ref{sigmapi_dr:fig}.
Finally, the vertex equation (\ref{Hedinvertex}) shows that vertex diagrams
arise from self-energy diagrams, by ``pinching'' a $G$ line with a vertex
(remove a line $G$ and replace it with $G\Gamma G$): this follows from the
functional definition of the vertex\cite{Hedin,Itzykson,Molinari05}.
In the GW approximation all corrections to the bare vertex are neglected:
the approximation $\Gamma =1$ replaces the exact vertex equation
(\ref{Hedinvertex}).

\section{Enumeration of Feynman diagrams} 
The perturbative solution of Hedin's equations in zero dimension provides
numerical coefficients which enumerate the Feynman diagrams for the five
correlators involved.
The ordinary perturbation expansion is carried out in the bare parameters
$v$ and $g$.
However, it is often convenient to expand in different ``renormalized''
variables, such as $v$ and $G$, or $W$ and $G$, or $G$,$W$ and $\Gamma$. 
These expansions count Feynman diagrams where, respectively, the propagator 
or both the propagator and the potential, or also the vertex, have been fully 
renormalized. 
Such diagrams where self-energy contributions of various types are resummed, 
are called {\sl skeleton} graphs\cite{Bjorken,Fetter}.
Different levels of resummations will be considered below, including a
renormalization based on GW correlators.

\subsection{$x=g^2v$-expansion}\label{x-scheme}
Standard perturbation theory in the bare interaction $v$ has been studied
in $d=0$ by various authors\cite{Cvi,Argy} by considering the path integral
formulation of the interacting field, which degenerates to an ordinary
integral. An approach based on Hedin's equations is very convenient to deal 
with the Hartree propagator $g$ in place of the bare one, and
explicit counting numbers can be obtained\cite{Molinari05}.

\begin{figure}
\centerline{\epsfig{file=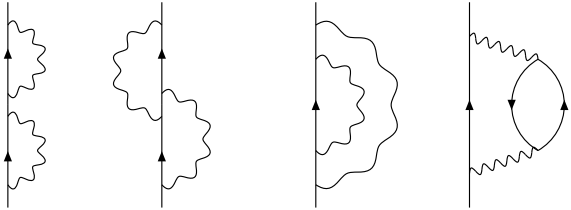,width=85mm,clip=}}
\caption{\label{perturb:fig}
The four diagrams for $G$ at second order in $x$.}
\end{figure}

Here, the natural dimensionless expansion parameter is $x=ig^2v$.
As an example, we provide the equation for $Z(x)=G/g$, where $Z$ is the
renormalization factor due to self-energy corrections. The algebraic
equations (\ref{HedinGW},\ref{Hedinsigmapi}) yield
\begin{eqnarray} \label{gammazeta}
  \ell xZ^2(x)\Gamma(x)= 1-\frac{1}{1-\ell +\ell Z(x)}
\,.
\end{eqnarray}
The GW approximation ($\Gamma=1$), generates a cubic equation for $Z_{GW}$.
For the complete theory, we rewrite the vertex equation
(\ref{Hedinvertex}), in a form\cite{Molinari05} where $G$ 
is traded for $g$:
\begin{eqnarray} \label{GammaZ}
\Gamma (x) = 1+ g^2 \frac{\partial\Sigma}{\partial g}\,=\,
\frac{1}{Z(x)} - 2x\frac{d}{dx}\frac{1}{Z(x)}
\,.
\end{eqnarray}
Equations (\ref{gammazeta}) and (\ref{GammaZ}) combine to a closed equation
for $Z$:
\begin{eqnarray} \label{Z}
2\ell x^2\frac{dZ}{dx} = 1-\ell xZ(x)
-\frac{1}{1-\ell+\ell Z(x)}
\,,
\end{eqnarray}
with the initial condition $Z(0)=1$. The solution as a series expansion in
$x$ provides the number of Feynman diagrams that contribute to each order
to the propagator $G$
\begin{eqnarray}\label{x-count}
Z(x) = 1 + x + (3+\ell)x^2 +(15+11\ell+\ell^2)x^3 +  \ldots 
\end{eqnarray}
The $\ell$ powers represent the loop contents.
For example, at second order in $x$, the theory involves 3 diagrams with no
loops, and 1 diagram with one loop; they are shown in Fig.~\ref{perturb:fig}.
By substituting the power series for $Z(x)$ into Eq.~(\ref{GammaZ}), we
obtain the diagram count for the vertex function:
\begin{eqnarray}
\Gamma(x) =   1 + x + 3\left( 2 + \ell \right) x^2 + 
   5\left( 10 + 9\,\ell + \ell^2 \right) x^3+  \ldots
\end{eqnarray} 
Table~\ref{diagram:tab} reports the total number of vertex diagrams up to
10$^{\rm th}$ order, obtained by taking $\ell=1$ in the expansion for
$\Gamma(x)$.

\begin{widetext}

\begin{table*}[t]
\caption{\label{diagram:tab}
The number of $n^{\rm th}$-order skeleton diagrams for the vertex function
in the five renormalization schemes considered in the text.
}
\begin{ruledtabular}
\begin{tabular}{crrrrrrrrrr}
$n$      	&	1	&	2	&	3	&	4	&	5	&	6	&	7	&	8	&	9	&	10	\\
\hline
$\Gamma(x)$	&	1	&	9	&	100	&	1323	&	20088	&	342430	&	6461208	&	133618275	&	3006094768	&	73139285178	\\
$\Gamma(y)$	&	1	&	7	&	63	&	729	&	10113	&	161935	&	2923135	&	58547761	&	1286468225	&	30747331223	\\
$\Gamma(t)$	&	1	&	6	&	52	&	602	&	8223	&	128917	&	2273716	&	44509914	&	957408649	&	22449011336	\\
$\Gamma(z)$	&	1	&	6	&	49	&	542	&	7278	&	113824	&	2017881	&	39842934	&	865391422	&	20486717908	\\
$\Gamma(u)$	&	1	&	3	&	13	&	147	&	1965	&	30979	&	559357	&	11289219	&	250794109	&	6066778627	\\
\end{tabular}
\end{ruledtabular}
\end{table*}

\end{widetext}


\subsection{$y=G^2v$-skeleton expansion}\label{y-scheme}
When all self-energy insertions of the propagator are resummed, one ends up
with Feynman diagrams where the propagator lines correspond to the exact
$G$, while the interaction lines continue to correspond to the bare
$v$. This resummation is obtained by using the expansion parameter
$y=iG^2v$.  It is possible to obtain a closed equation for the vertex
function. First write the self-energy as
\begin{eqnarray} \label{sigmay}
\Sigma = iGW\Gamma = iG\frac{v}{1-v\Pi}\Gamma =
\frac{1}{G} \frac{y\Gamma}{1-y \ell \Gamma}
\,.
\end{eqnarray}
Next evaluate the vertex function in Eq.~(\ref{Hedinvertex}) by means of
Eq.~(\ref{sigmay}),
and obtain a differential equation for $\Gamma(y)$:
\begin{eqnarray}\label{gammay}
 2y^2\Gamma \frac{d\Gamma}{dy}
&=&-1+(1+2y\ell )\Gamma -y\Gamma^2(1+2\ell +y\ell^2)\nonumber \\ 
&&+ y^2\Gamma^3 \ell(\ell-1)
\,.
\end{eqnarray}
This equation must be solved with the initial condition $\Gamma (0)=1$.
The series expansion of the solution counts skeleton vertex diagrams
with dressed propagators and bare interactions:
\begin{eqnarray}\label{y-count}
\Gamma (y) = 1+y+(4+3\ell)y^2+(27+31\ell+5\ell^2 )y^3+\ldots
\end{eqnarray}
Table~\ref{diagram:tab} lists the total number of these diagrams up to
order $n=10$, by taking $\ell=1$ in the expansion for $\Gamma(y)$, as
was done above.

The enumeration of skeleton diagrams of the polarization
$\Pi=i\ell G^2\Gamma (y)$ coincides with that of $\Gamma (y)$.
The skeleton expansion of $\Sigma$ results from Eq.~(\ref{sigmay}):
\begin{eqnarray}
\Sigma/iGv = 1+ (1+\ell)y +  (4+5\ell+\ell^2)y^2\nonumber \\
+(27+40\ell+14\ell^2+\ell^3)y^3+\ldots
\end{eqnarray}
In the expansion in terms of the bare parameter $x$, $\Pi$ and $\Sigma$
shared the diagram counting,\cite{Molinari05} while this property does not
apply to the $y$ expansion at hand.

The $G^2v$-expansion of $\Sigma $ is interesting for the theory of the 
Luttinger-Ward $\Phi $-functional\cite{Almbladh99,Vignale05}.
By closing all $\Sigma [G,v]$ skeleton graphs with a $G$ line, and dividing
each one by the number of $G$ lines it contains, one obtains a functional
with the property of yielding the self-energy and the reducible
polarization:
\begin{eqnarray}\label{PhiDefinition}
\Sigma(1,2)= \frac{\delta\Phi[G,v]}{\delta G(2,1)},\qquad \tilde\Pi (1,2)=
-2\frac{\delta \Phi[G,v]}{\delta v(2,1)}
\,.
\end{eqnarray}
In $d=0$, the functional $\Phi[G,v]$ turns into an ordinary function of
$y=iG^2v$.
\begin{eqnarray}\nonumber
\Phi(y) &=& \frac{1}{2}\int_0^y dy' \frac{\Gamma (y')}{1-\ell y'\Gamma (y')}
\\
&=&\frac{y}{2}+ (1+\ell)\frac{y^2}{4}+(4+4\ell+\ell^2)\frac{y^3}{6}+
\ldots 
\end{eqnarray}
This last expression provides the appropriate diagram counting and
fractional weights for $\Phi(y)$ in any dimension.

\subsection{$z=G^2 W$-skeleton expansion}\label{z-scheme}
Resummation of both self-energy and polarization insertions leads to
skeleton diagrams with exact propagators $G$ and interactions $W$.
Here, the natural expansion parameter is $z=iG^2W$.
The parameter $z$ is linked as follows
\begin{eqnarray}
z =iG^2\frac{v}{1-i\ell v\Gamma G^2}=\frac{y}{1-\ell y\Gamma } 
\end{eqnarray}
to the parameter $y=iG^2v$ of the previous expansion (\ref{y-count}).
Inversion yields $y=z\,[1+\ell z\Gamma (z)]^{-1}$.
By entering this relation into the differential equation (\ref{gammay})
for $\Gamma (y)$, which is now viewed as a function of $z$, we obtain
\begin{eqnarray}\label{gammaz}
z^2\frac{d\Gamma (z)}{dz} = \frac{1 - \Gamma(z) +z\Gamma^2(z) +
2\ell z^2\Gamma^3(z)}{\ell-(2+\ell)\Gamma (z)- 3z\ell \Gamma^2(z)} 
\,.
\end{eqnarray}
The appropriate series expansion counts $G^2W$-dressed vertex diagrams:
\begin{eqnarray}\label{z-count}
\Gamma (z) = 1 + z + 2\left(2+\ell\right)z^2 + \left(27+22 \ell\right) z^3
+\ldots
\end{eqnarray}
Table~\ref{diagram:tab} lists the total number of these $\Gamma(z)$
diagrams up to order $n=10$.
Both $\Sigma =iGW\Gamma (z)$ and $\Pi=i\ell G^2 \Gamma (z)$ 
have the same total $G^2W$-counting numbers as the vertex.

In the $G^2W$-expansion, it is natural to define \cite{Almbladh99} the
functional $\Psi[G,W]$, which generalizes Eq.~(\ref{PhiDefinition}) and
generates the self-energy and the irreducible polarization as follows:
\begin{eqnarray}\label{PsiDefinition}
\Sigma(1,2)= \frac{\delta\Psi[G,W]}{\delta G(2,1)},\qquad \Pi (1,2)=
-2\frac{\delta \Psi[G,W]}{\delta W(2,1)}
\,.
\end{eqnarray}
The diagrammatic construction of $\Psi $ is the same as for $\Phi$: close
the skeleton graphs of $\Sigma =iGW\Gamma[G,W] $ with a $G$ line and divide
by the number of $G$ lines that the graph contains.
In $d=0$ we have $G\Sigma =z\Gamma(z)$; to divide by the
number of $G$ lines that each $G\Sigma$-diagram contains, corresponds to 
the integral 
\begin{eqnarray}
\Psi(z)=\frac{1}{2}\int_0^z dz' \Gamma(z')= \frac{z}{2}+\frac{z^2}{4}+
(2+\ell)\frac{z^3}{3}+\ldots
\end{eqnarray}

\subsection{$u=G^2W\Gamma^2$-skeleton expansion}\label{u-scheme}
The $G^2W-$expansion resums all self-energy and polarization insertions,
and skeleton diagrams of various order in $z$ result just because of vertex
contributions.
If vertex diagrams are summed as well, each of the self-energy and the
polarization is brought to a unique skeleton diagram depicted in
Fig.~\ref{sigmapi_dr:fig}.
Vertex diagrams cannot
be brought to a finite collection of skeleton diagrams. 
We may nevertheless resum vertex insertions in all vertex diagrams, 
and enumerate the resulting vertex skeleton diagrams, where all lines and 
vertices are resummed.

\begin{figure}
\centerline{\epsfig{file=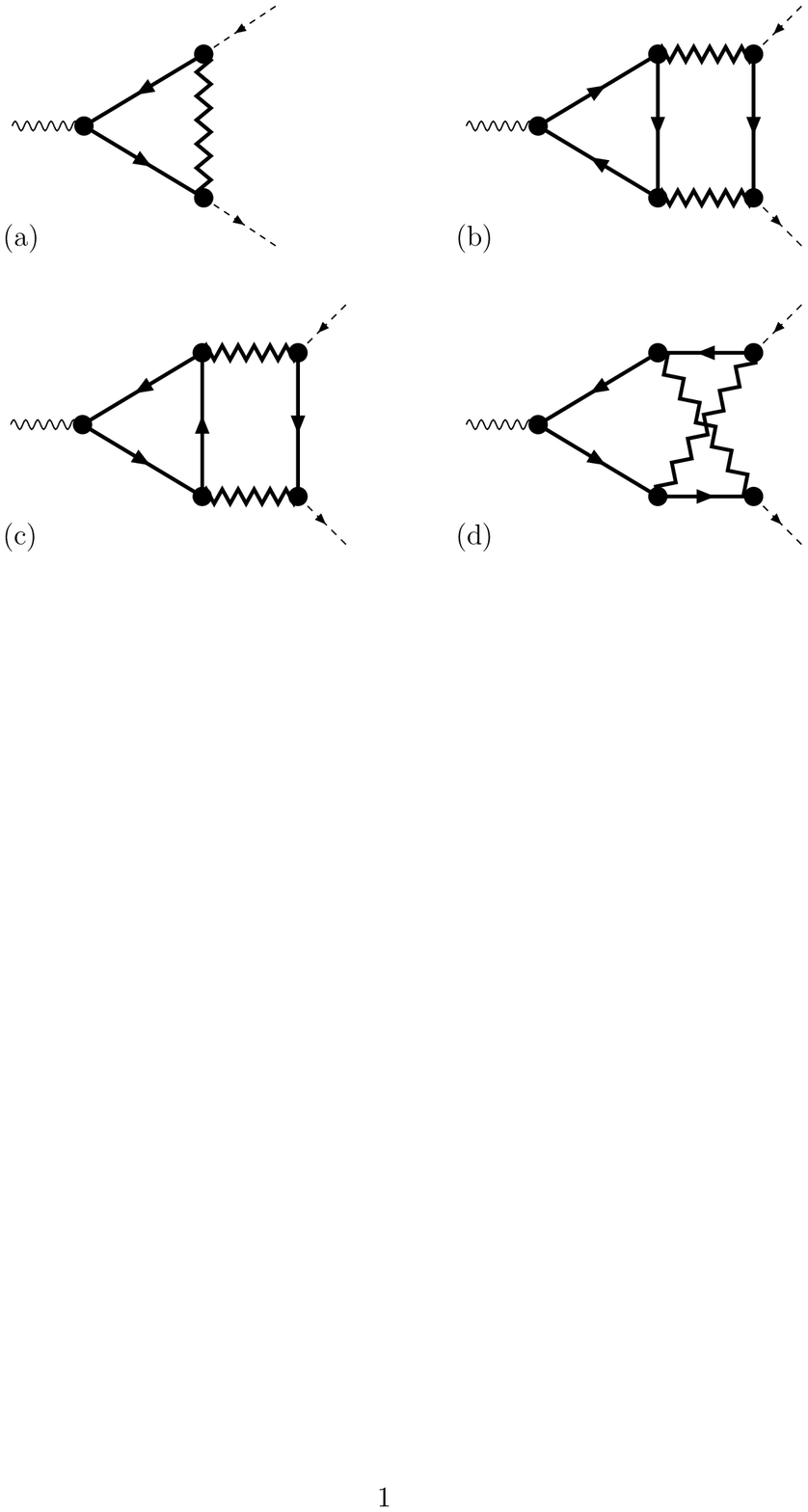,width=85mm,clip=}}
\caption{\label{vertex_dr:fig} All first-order (a) and second-order (b,c,d) vertex 
diagrams in $u$ expansion.}
\end{figure}

This is achieved by noting that in a vertex diagram, each $W$ line ends in two
vertices. However, the vertex which the Coulomb external line is attached to, 
is left out (see Fig.~\ref{vertex_dr:fig} for some vertex diagrams).
We then write $\Gamma = 1+\Gamma \gamma (u)$, in terms of the expansion
variable $u=iG^2W\Gamma^2=z\Gamma^2(z)$.  The factor $\Gamma$ in front of
$\gamma (u)$ accounts for the left-out vertex.

The Taylor expansion of $\gamma (u)$ enumerates all dressed fully 
renormalized vertex diagrams. An equation for $\gamma $ is obtained
from the relation
\begin{eqnarray}
\frac{d\gamma}{du}=\frac{dz}{du}\frac{d}{dz}\frac{\Gamma -1}{\Gamma}=
\frac{1}{\Gamma ^2}\frac{d\Gamma/dz}{\Gamma^2+2z\Gamma d\Gamma/dz}
\,,
\end{eqnarray}
where we enter $d\Gamma/dz$ given by Eq.~(\ref{gammaz}):
\begin{eqnarray}
u \frac{d\gamma}{du}=(1-\gamma)
\frac{2\ell u^2(1-\gamma)^2+u(1-\gamma)-\gamma }
{\ell u^2 (1-\gamma)^2 -\ell u \gamma (1-\gamma) -2\gamma}
\,.
\end{eqnarray}
This equation is solved with initial condition $\gamma (0)=0$.
The resulting $u$-expansion for $\Gamma$ is
\begin{eqnarray}\nonumber
\Gamma(u)&=& 1 + (\Gamma u) + (1+2\ell)(\Gamma u^2) + (7+6\ell)(\Gamma u^3)\\
&&+(63+74\ell+10\ell^2)(\Gamma u^4) +\ldots 
\label{u-count}
\end{eqnarray}
The diagrams of first and second order are shown in
Fig.~\ref{vertex_dr:fig} (the next 13 third-order diagrams coincide with
those of QED, drawn in Ref.~\onlinecite{Cvi}).
This $u$-expansion represents the ``ultimate'' skeleton expansion, where all
ingredients of Hedin's equations have been renormalized.
Indeed, as apparent in Table~\ref{diagram:tab}, the diagram count is
smallest in the $u$ expansion at hand.

\subsection{$t=(G^2 W)_{GW}$-skeleton expansion}\label{t-scheme}
In the GW approximation ($\Gamma=1$), in physical dimensions, Hedin's
equations are a system of ordinary integral equations for $G_{GW}$,
$\Sigma_{GW}$, $W_{GW}$ and $\Pi_{GW}$.
It is conceivable that in a near future increased computer power will allow
us to solve these equations in fully self-consistent GW: this would make
the GW approximation the zeroth-order stage of a subsequent attack of the
full many-body problem, where vertex corrections are included
perturbatively.
The problem of including vertex corrections in a systematic way is an
important target of present-day research, to overcome several limitations
of the non-conservative GW \cite{Schindlmayr98,Oni,Bruneval05}.
It is therefore of interest to count the diagrams where $g$ and $v$ lines
are dressed (resummed) to include all their GW self-energy and polarization
insertions.

The counting problem is solved in $d=0$ by considering the expansion
parameter $t=iG^2_{GW}W_{GW}$. With simple algebra, the following two
equations for the vertex and the ratio $Z_{GW}(t)=G/G_{GW}$ are obtained:
\begin{eqnarray}
t Z_{GW}^2 (t)\Gamma (t)&=&\frac{[(1+t)Z_{GW}(t)-1](1+\ell t)}{1-\ell +\ell 
(1+t)Z_{GW}(t)}
\,,\\
2t^2 \frac{dZ_{GW}}{dt}&=&
(1-t-2\ell t^2)\frac{(1+t)Z_{GW}(t)-1}{1-
\ell +\ell (1+t)Z_{GW}(t)}\nonumber\\
&&-\frac {t}{1+\ell t}\, Z_{GW}(t)
\,.
\end{eqnarray}
The perturbative solutions are:
\begin{eqnarray}
 Z_{GW}(t)&=&1 + t^2 + (5+3\ell)t^3 +\ldots
\\\label{t-count}
\Gamma (t)&=&1 + t+ 2 (2+\ell)t^2 +(29 + 23 \ell)t^3 +\ldots
\end{eqnarray}
Table~\ref{diagram:tab} lists the total number of these $\Gamma(t)$
diagrams up to order $n=10$.

\section{Asymptotics}


\begin{figure}
\centerline{\epsfig{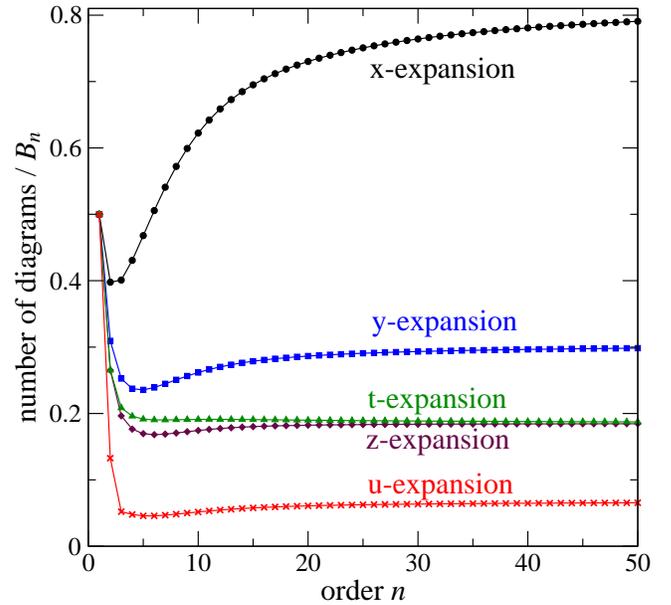}}
\caption{\label{asy:fig} (color online)
Number of vertex diagrams at order $n$, divided by the asymptotic
expression $B_n$ defined in Eq.~(\ref{asy:eq}), for the different
renormalization schemes realized in
Sec.~\ref{x-scheme} [$x$ -- Eq.~(\ref{x-count})],
Sec.~\ref{y-scheme} [$y$ -- Eq.~(\ref{y-count})],
Sec.~\ref{t-scheme} [$t$ -- Eq.~(\ref{t-count})],
Sec.~\ref{z-scheme} [$z$ -- Eq.~(\ref{z-count})], and
Sec.~\ref{u-scheme} [$u$ -- Eq.~(\ref{u-count})].
}
\end{figure}

To estimate asymptotic behaviors is not simple in the present approach.
However, the topology of many-body diagrams with two-body interaction is
the same as in relativistic quantum electrodynamics (QED), with a
difference: because of exact particle-antiparticle symmetry, QED diagrams
with loops involving an odd number of fermion lines cancel (Furry's
theorem \cite{Itzykson}). Therefore, many-body skeleton diagrams grow
faster in number than QED ones.

The functional approach, in the saddle point expansion, is better
suited for asymptotics.
This problem in physical dimensions has been studied for QED by several
authors, based on Lipatov's powerful method \cite{Itzykson77,Suslov05}.
%
%
%
From the tabulated QED asymptotic values \cite{Cvi} in $d=0$, one can infer
the leading behavior of many-body vertex skeleton diagrams.
Consider the series for the vertex $\Gamma(\theta) = \sum_n A_n\theta^n$,
in one of the renormalization schemes ($\theta =x,\ y,\ z,\ t,\ u$)
outlined above.  We obtain
\begin{eqnarray}\label{asy:eq}
 A_n = B_n\, (c_0+c_1/n +\ldots )
\,, \qquad
B_n=n!\, 2^n\, n^{3/2}\,.
\end{eqnarray}
In Fig.~\ref{asy:fig} the asymptotic behavior is checked on the first 50
coefficients of the five different vertex skeleton expansions computed in
the present work.
Indeed, in all expansions, whether bare or renormalized, the number of
vertex diagrams grows with the order $n$ roughly as $B_n$, thus leading to
divergent power series, like in QED.
The renormalization scheme only affects the subleading coefficients $c_i$.
As expected, the ``ultimate'' $u$ expansion, involving wider diagram
resummations leads to fewer diagrams than all other ``less renormalized''
expansions.
For large $n$, the diagram count is essentially the same for the $z$
expansion, including renormalization to both the propagator and the
interaction, and for the $t$ expansion around Hedin's GW
approximation.

\section{Conclusions}

The present method for enumerating skeleton diagrams is based on Hedin's
equations and it is very efficient.
Simple changes of the expansion variable produce various renormalizations
schemes, in the form of differential equations that can easily be solved by
series.
The coefficients of these series are integers that count skeleton diagrams.
These are useful as a check for any diagrammatic approach to the many-body
problem in realistic dimensions.
As expected, the number of diagrams, whether renormalized or not, grows
factorially with the order.

\begin{acknowledgments}
This work was funded in part by the EU's 6th Framework Programme through 
 the NANOQUANTA Network of Excellence (NMP4-CT-2004-500198)
\end{acknowledgments}



\end{document}